\documentclass[aip,reprint]{revtex4-1}
\usepackage{graphicx}
\usepackage{placeins}
\usepackage{float}

\begin{document}


\title{High-Throughput Microwave Package for \\Precise Superconducting Device Measurement} 



\author{Wei-Ren Syong}
\affiliation{Department of Physics, University of Colorado, Boulder, CO 80309, USA}
\affiliation{Department of Electrical, Computer, and Energy Engineering, University of Colorado, Boulder, CO 80309, USA}
\affiliation{National Institute of Standards and Technology, Boulder, CO 80305, USA}
\author{Allie Miller}
\affiliation{Department of Electrical, Computer, and Energy Engineering, University of Colorado, Boulder, CO 80309, USA}
\author{Emma Davis}
\affiliation{Department of Mechanical Engineering, University of Colorado, Boulder, CO 80309, USA}
\author{John R. Pitten}
\affiliation{Department of Physics, University of Colorado, Boulder, CO 80309, USA}
\affiliation{Department of Electrical, Computer, and Energy Engineering, University of Colorado, Boulder, CO 80309, USA}
\affiliation{National Institute of Standards and Technology, Boulder, CO 80305, USA}
\author{Jorge Ramirez}
\affiliation{Department of Physics, University of Colorado, Boulder, CO 80309, USA}
\affiliation{Department of Electrical, Computer, and Energy Engineering, University of Colorado, Boulder, CO 80309, USA}
\affiliation{National Institute of Standards and Technology, Boulder, CO 80305, USA}
\author{Nathan Ortiz}
\affiliation{National Institute of Standards and Technology, Boulder, CO 80305, USA}
\author{Michael Vissers}
\affiliation{National Institute of Standards and Technology, Boulder, CO 80305, USA}
\author{Doug Bennett}
\affiliation{National Institute of Standards and Technology, Boulder, CO 80305, USA}
\author{Corey~Rae Harrington McRae}
\affiliation{Department of Electrical, Computer, and Energy Engineering, University of Colorado, Boulder, CO 80309, USA}
\affiliation{Department of Physics, University of Colorado, Boulder, CO 80309, USA}
\affiliation{National Institute of Standards and Technology, Boulder, CO 80305, USA}

\date{\today}

\begin{abstract}
Cryogenic microwave measurement of superconducting quantum devices is complicated by the packaging required to connect devices to control and readout circuitry. In this work, we outline the design and experimental demonstration of a wirebond-free, PCB-free, drop-in microwave package for on-chip superconducting quantum devices. The package is composed of a superconducting aluminum cavity with a suspended tungsten transmission pin. The fundamental package cavity mode is far detuned from the 4 GHz to 8 GHz band of interest, and the pin transmission exhibits less than 3 dB of ripple across this range. We demonstrate the use of this package to extract the loss tangent of superconducting ring resonators, measuring a value of $(1.10 \pm 0.09) \times 10^{-6}$, which agrees with measurements of $\lambda/4$ resonators in wirebond-based packaging. This high-throughput measurement system will allow the rapid generation of large datasets for improving superconducting qubit performance, and facilitate time-sensitive surface passivation and oxide regrowth studies.
\end{abstract}


\maketitle 


High-performance superconducting qubit relaxation times are limited by the dielectric loss tangent of qubit component materials.~\cite{Muller2019,Read2023,Bal2024} Low-power measurements of planar superconducting resonators are commonly used to precisely characterize this loss.~\cite{McRae2020} Conventional on-chip superconducting quantum devices must be secured in a package in order to connect the chip to control and readout lines and install the chip in a cryogenic environment such as a dilution refrigerator. Commonly, a machined normal or superconducting metal box is used to house the chip, which is wirebonded to a printed circuit board (PCB) for signal routing.

However, these packages have several performance pitfalls. Sample installation can be time-consuming, making time-sensitive interface studies difficult and reducing the throughput of cryogenic microwave measurement. Additionally, packaging will add microwave nonidealities such as Fano interference, spurious modes, and insertion loss.~\cite{Khalil2012,Rieger2023,Pitten2025,Wenner2011,Bronn2018,McConkey2018,McRae2017}

The measurement of many devices is usually needed to accurately characterize loss due to resonator-to-resonator variation.~\cite{Woods2019} As resonator measurement time can be reduced by many methods recently adopted by the community, including homophasal~\cite{Baity2024} and multiplexed measurement, packaging time is threatening to bottleneck the measurement process. 

Separately, the time devices spend in air can be critical to an experiment. Surface oxides are of significant interest in superconducting qubit studies due to their outsized impact on qubit relaxation.~\cite{Place2021,Verjauw2021,Bal2024} These oxides can be removed with fluorine-based etching but regrow with air exposure. For this reason, surface passivation techniques are an active area of study. These experiments require rapid packaging in order to precisely vary the device time spent in air.

\begin{figure}
\includegraphics[width=1.0\linewidth]{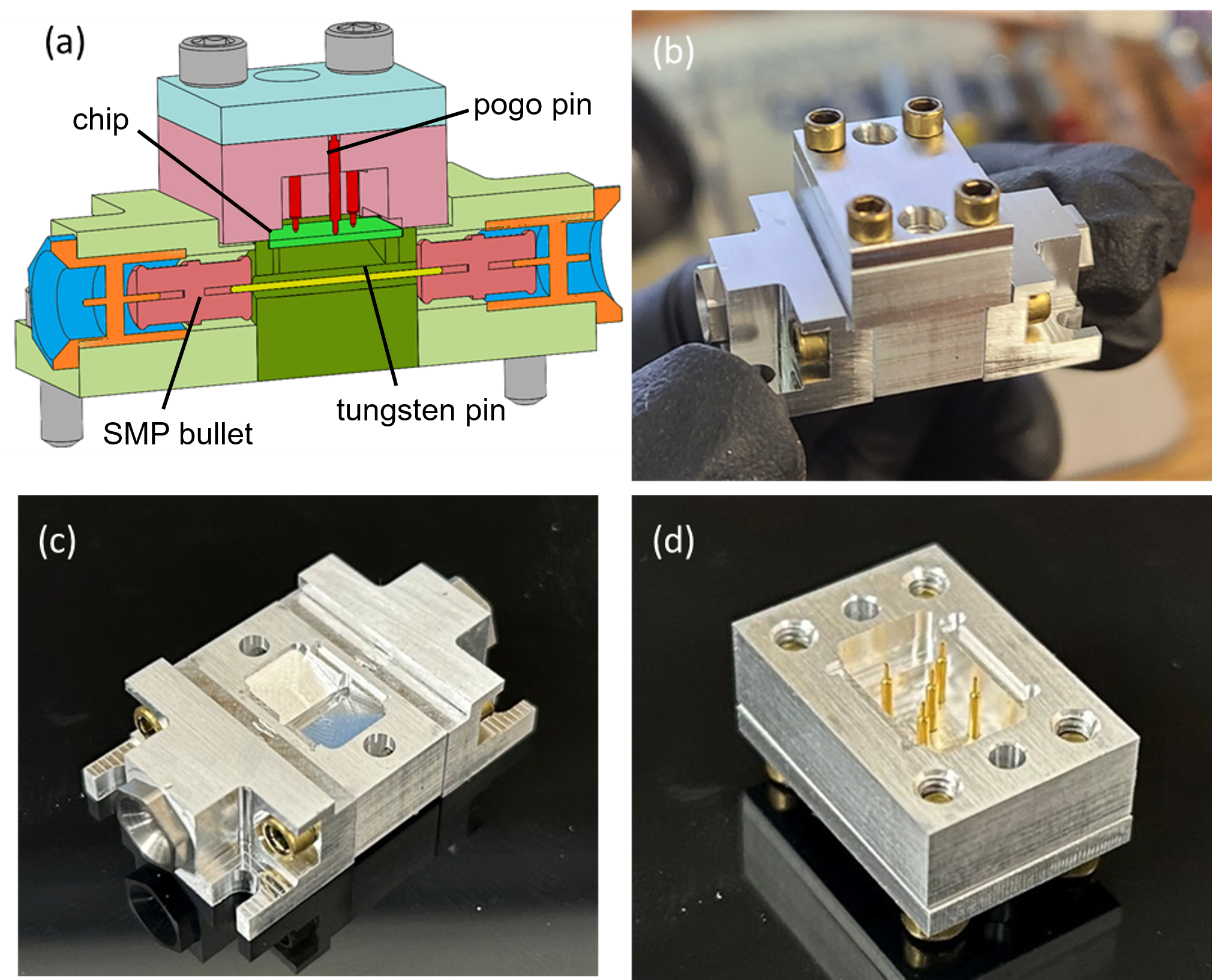}
\caption{Microwave package design. (a) CAD model showing tungsten pin (yellow), pogo pins (red), and installed chip (bright green). The package consists of four main parts: the cavity base (dark green), the lid (light blue), the interposer (pink), and the SMP holders (pale green). The tungsten pin feeds directly into SMP bullets (dark pink) on either side. (b)-(d) Photographs of closed package (b), cavity base and SMP holders (c), and lid and interposer (d).
\label{fig:packagedesign}}
\end{figure}

Accurate and precise extraction of device parameters such as resonance frequency, quality factor, and dielectric loss tangent requires a clean microwave background with few environmental modes or impedance mismatches, and low frequency-dependent insertion loss. Wirebonds in particular can lead to impedance mismatch and thus Fano interference in on-chip superconducting resonators.~\cite{Khalil2012,Rieger2023,Pitten2025}

In this work, we demonstrate a transmission-pin-coupled superconducting package with a clean microwave background designed for rapid chip swap-out. We use this package to measure a set of superconducting ring resonators and demonstrate consistent two-level-system (TLS) loss tangent extraction with a small standard deviation compared to $\lambda / 4$ coplanar waveguide resonators measured using a standard package.

The sample package is installed at the mixing chamber stage of a dilution refrigerator with a base temperature of approximately 15~mK. The input microwave lines are attenuated by a total of 70 dB, distributed across multiple temperature stages. The output signal is amplified first by a high-electron-mobility transistor (HEMT) amplifier located at the 4 K stage and then by a room-temperature amplifier before being received by the vector network analyzer (VNA). The cavity is installed between two six-way cryogenic microwave switches, along with a $\lambda / 4$ resonator in a wirebonded package with soldered PCB, as well as a through line for inter-cooldown comparison. Additional details are provided in the Supplementary Material.

The package design is shown in Fig.~\ref{fig:packagedesign}. It consists of a machined 6061 aluminum cavity with a suspended tungsten transmission pin that connects two non-magnetic SMP connectors. No PCBs or other bulk dielectric materials are present in the cavity. The tungsten pin is positioned within a rectangular cutout below the main cavity, forming a coax-like geometry whose characteristic impedance, estimated using the standard coaxial waveguide formula, is close to 50 $\Omega$. Additional cavity design details are described in the Supplementary Material. 

\begin{figure}
\includegraphics[width=\linewidth]{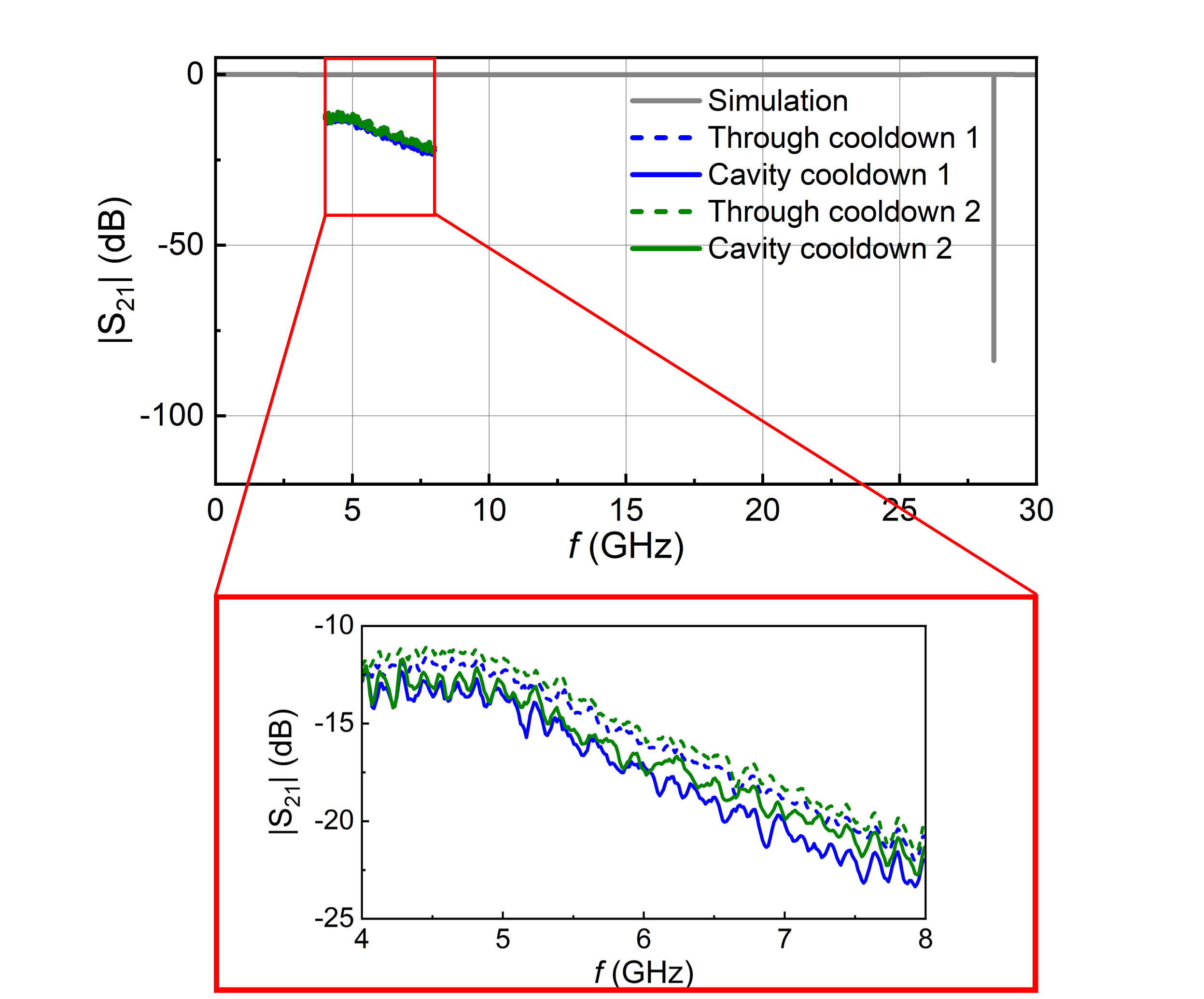}%
\caption{Experimental and simulated transmission through the microwave package. HFSS simulation of the empty package is shown (solid grey). Cryogenic measurements of empty package (blue) and associated through (dashed blue) in cooldown 1, and package with ring resonator chip installed (green), and associated through (dashed green) in cooldown 2 are shown. Comparison between package and through is facilitated by the use of cryogenic switches. The close agreement between package and through traces demonstrates a clean microwave background in the cavity package.
\label{fig:cavityressim}
}%
\end{figure}

The fundamental cavity mode is designed to be well above the 4~GHz to 8~GHz operational frequency range of the on-chip superconducting resonators. Although the cavity modes are not used for readout, they are simulated to ensure a low-ripple microwave background within the measurement band. As shown in Fig.~\ref{fig:cavityressim}, Ansys HFSS simulations predict the fundamental mode near 28~GHz, with background ripple across the 4 GHz to 8 GHz of less than 3~dB. The details are shown in Supplementary Material. From this, we expect expect that this package could be used for devices at lower and higher frequencies than 4 GHz to 8 GHz, but we are limited to this bandwidth by our measurement chain in this experiment.

To experimentally verify these simulations, transmission through the empty cavity is measured and compared to a reference through (Fig.~\ref{fig:cavityressim} inset) using cryogenic switches to swap between devices. While the absolute transmission level in the measurements is reduced relative to simulation due to insertion loss from the measurement chain, the cavity and through traces exhibit qualitatively similar frequency dependence and texture across the 4 GHz to 8~GHz bandwidth of interest. The difference between the cavity and through responses remains below 3~dB, indicating that the cavity itself introduces minimal additional attenuation or ripple beyond the system background. As a result, the microwave background is sufficiently small and slowly varying to be effectively frequency independent over the narrow linewidth of each resonance. Under these conditions, the diameter-corrected circle fit can be applied directly without requiring additional assumptions about background frequency dependence, enabling accurate extraction of $Q_i$. Additional comparisons are provided in the Supplementary Material. 

\begin{figure}
\includegraphics[width=\linewidth]{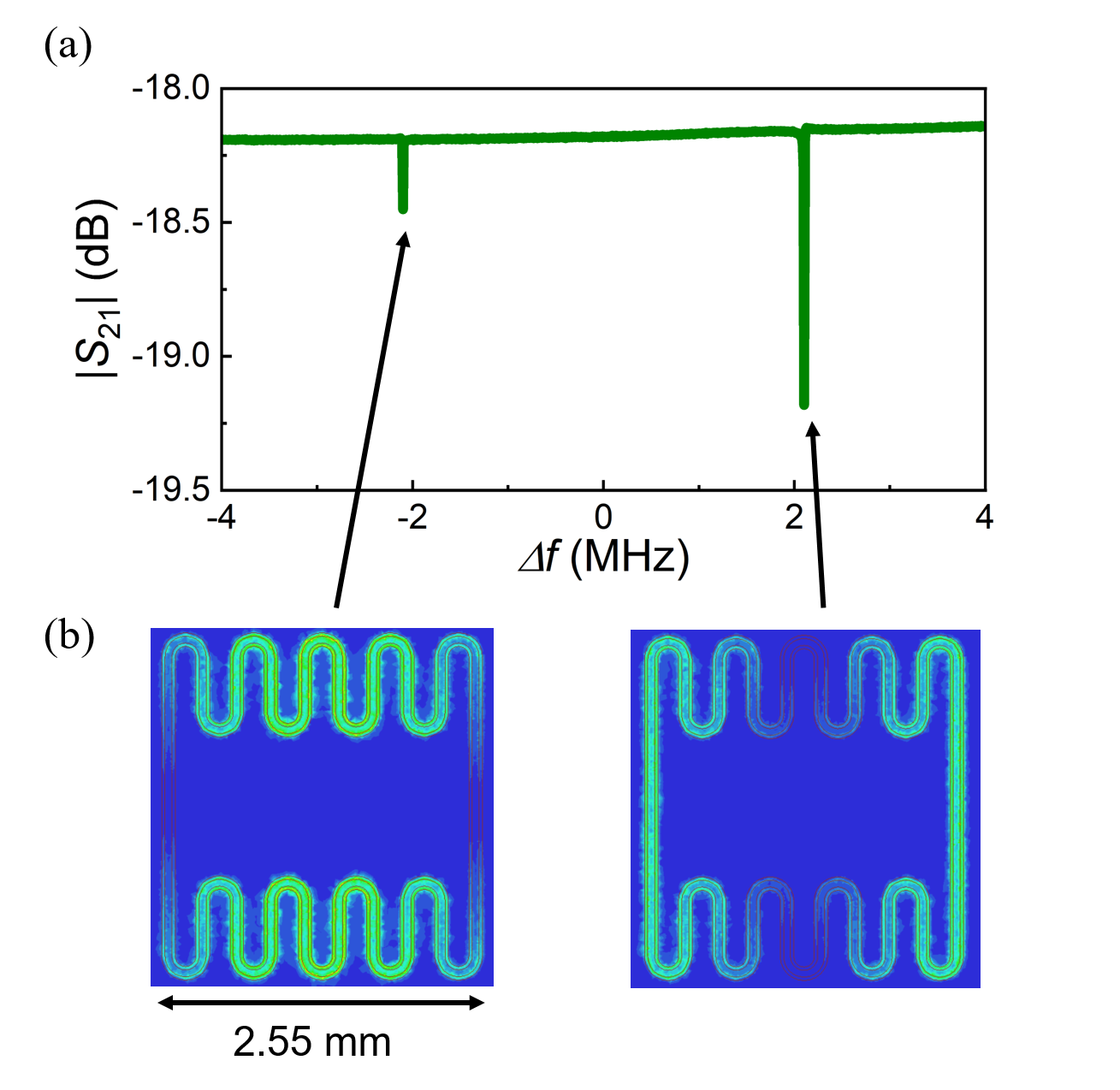}%
\caption{Characterization of a ring resonator using the cavity package. (a) Measured $S_{21}$ transmission showing the mode splitting for one ring resonator. (b) Ansys HFSS simulation showing the electric field distributions of the two standing-wave modes of a single ring resonator.
\label{fig:ringressplit}}%
\end{figure}

To demonstrate the functionality of the high-throughput microwave package, we fabricated and measured a set of on-chip superconducting ring resonators (Fig.~\ref{fig:ringressplit}(a)). Devices are patterned on a double-side-polished (DSP), float-zone intrinsic Si substrate with a thickness of 380~$\mu$m. The DSP nature of the substrate eliminates the possibility of additional loss due to wafer etching. Prior to deposition, the substrate is treated with an hydrofluoric acid vapor clean to remove native oxides. A 200~nm Nb film is then sputter-deposited and patterned using i-line photolithography, followed by an SF$_6$-based reactive ion etch. Finally, a 3 minute O$_2$ plasma clean is performed to remove resist residues and surface contaminants. Stress measurement shows around 50~MPa of compressive stress in the Nb film.

Each 7.5~mm $\times$ 7.5~mm chip contains four ring resonators. No on-chip feedline is included, as signal readout is performed using the suspended tungsten transmission pin. The ring resonators are meandered, closed, full-wavelength coplanar waveguide (CPW) loops, with a center conductor width of 50~$\mu$m and a ground gap of 25~$\mu$m. Ring perimeters range from 17~mm to 21~mm, corresponding to resonance frequencies within the 4 GHz to 8~GHz band. The additional ring resonator details are shown in the Supplementary Material.

Each chip is installed with the patterned film facing the tungsten pin. The chip is mechanically secured using non-magnetic pogo pins placed between the package lid and the backside of the chip, as shown in Fig.~\ref{fig:packagedesign}(a). This approach avoids the use of adhesives, pastes, or cryogenic-compatible paints, which can introduce unwanted loss or contamination.

As shown in Fig.~\ref{fig:cavityressim}, pin transmission in the presence of the chip is similar to that of the empty cavity when compared to their associated throughs. Resonance dips are not visible in Fig.~\ref{fig:cavityressim} because the broadband sweep uses a coarse frequency spacing relative to the narrow linewidths of the resonator modes, which undersamples the resonances. The observed agreement between the chip-loaded and empty cavity responses confirms that the overall transmission environment remains clean and reproducible with chip installation.

\begin{figure}[H]
\includegraphics[width=1.0\linewidth]{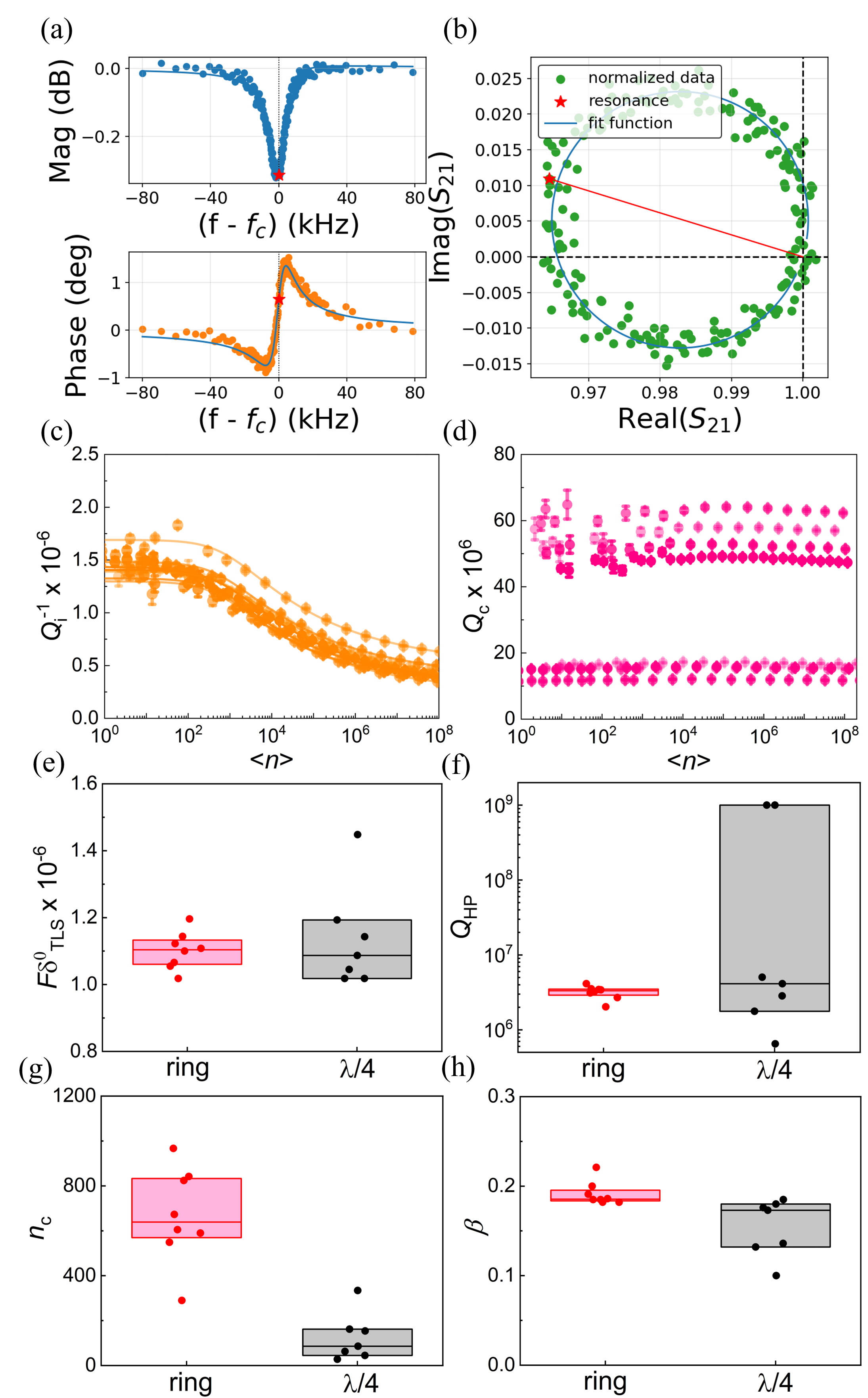}%
\caption{Microwave characterization of ring resonators in the superconducting cavity package. (a) and (b) Diameter correction method (DCM) fit of a single resonance in the low-power regime. A ring resonator measured at an estimated on-chip power of -136~dBm ($\langle n \rangle = 4$) is shown as an example. Points are normalized data, star is resonance, and line is fit to DCM. (c) and (d) Inverse internal quality factor $Q_i^{-1}$   and coupling quality factor $Q_c$ as a function of average number of photons $\langle n \rangle$ for all resonances. Error bars denote the 95~$\%$ confidence interval for each circle fit, and lines denote fits to Eq.~\ref{eq:scurve}. (e)-(h) Extracted loss parameters for the ring resonators in the cavity package compared to $\lambda/4$ resonators measured in a standard wirebonded package. TLS loss tangent $F \delta^0_{\mathrm{TLS}}$, high power quality factor $Q_{\mathrm{HP}}$, critical photon number $n_c$, and empirical parameter describing TLS interaction $\beta$ are compared. Box shows the median value and the 25th and 75th percentiles, and points denote individual measurements.
}
\label{fig:ringresmeas}%
\end{figure}

Slotline modes, which often arise in CPW geometries due to unequal ground-plane potentials,~\cite{Pozar} are neither observed in experimental measurements nor significantly coupled in simulation (see Supplementary Materials for details). Ansys HFSS simulations indicate that the lowest-order slotline mode occurs near 10~GHz, well above our 4-8~GHz operational range, and is weakly coupled to the measurement circuit. This eliminates the need for on-chip wirebonds, and we do not use them in this study.

A typical measured $S_{21}$ response of a single ring resonator is shown in Fig.~\ref{fig:ringressplit}(a). In the ideal case of perfect symmetry, each ring supports two degenerate standing-wave modes that share the same resonance frequency. In practice, this degeneracy is lifted by impedance perturbations along the resonator, resulting in a doublet with slightly different resonance frequencies, as described in Ref.~\cite{Sun2025}. In that framework, such perturbations arise from fabrication imperfections and spatial inhomogeneities in the superconducting film and surrounding dielectric. In the present devices, the observed frequency splittings are smaller than those reported in Ref.~\cite{Sun2025}, which may reflect improved fabrication uniformity and the larger resonator size used here. Additionally, the asymmetric coupling geometry associated with the transmission pin may provide an additional source of symmetry breaking in our implementation, although its quantitative contribution is not isolated in this work. The corresponding simulated electric field distributions for the two modes are shown in Fig.~\ref{fig:ringressplit}(b).   

The coupling between the on-chip resonators and the suspended transmission pin follows a similar design approach to that described in Ref.~\citenum{Ganjam2024}. In this configuration, the pin acts as a weakly coupled probe. While the resonator geometry does influence the coupling strength, it can also be externally tuned by varying the vertical spacing between the pin and the chip. This approach provides experimental flexibility, as the same chip can be used with different package configurations to achieve the desired coupling regime, unlike devices with on-chip transmission lines which require a new chip with a different design in order to modify coupling strength. Details are provided in the Supplementary Material.

For each resonance, the raw, homophasal $S_{21}$ data are analyzed using a diameter-corrected circle fit,~\cite{Khalil2012} which extracts key parameters including the internal quality factor ($Q_i$), coupling quality factor ($Q_c$), and resonance frequency ($f_c$). An example of this fit is shown in Fig.~\ref{fig:ringresmeas}(a) and (b), corresponding to an estimated on-chip photon number $\langle n \rangle \sim 4$. At such low photon numbers, the signal-to-noise ratio is reduced, and longer acquisition times and increased averaging are therefore used. The TLS loss tangent is extracted from a global fit across multiple powers and is thus insensitive to the noise present in any single low-power trace.

$Q_c$ values for all ring resonators at all measured powers are reported in Fig.~\ref{fig:ringresmeas}(d), showing good agreement with HFSS simulations presented in the Supplementary Material. The band of $Q_c$ values around $10^7$ ($5 \times 10^7$) are linked to the split mode with the higher (lower) resonance frequency. The difference in $Q_c$ between the two modes can be attributed to different degrees of mode rotation, which results in variations in the electric field orientation and therefore in the strength of coupling to the transmission pin.~\cite{Sun2025} While the assignment of a given coupling strength to a particular standing-wave mode depends on the chosen phase convention, the observed disparity in $Q_c$ between the two split modes reflects their rotated field profiles relative to the coupler.

The intrinsic dielectric losses of the ring resonators are characterized by measuring $Q_i$ as a function of input powers. At low excitation powers, loss is dominated by two-level-system (TLS) defects in surface oxides and material interfaces.~\cite{GaoThesis,Pappas2011,Muller2019} As excitation power increases, TLS become saturated, and the dominant loss mechanisms transition to power-independent processes such as quasiparticle generation or radiation loss.

Power-dependent loss data are fit to the S-curve model:
\begin{equation}
    \frac{1}{Q_i} = F \delta_{\mathrm{TLS}}^0 \frac{\tanh\left(\frac{\hbar \omega}{2 k_B T}\right)}{\left(1+\langle n \rangle / n_c\right)^{\beta}} + \frac{1}{Q_{\mathrm{HP}}},
    \label{eq:scurve}
\end{equation}
where $F$ is the filling factor, $\delta_{\mathrm{TLS}}^0$ is the intrinsic TLS loss tangent, $\langle n \rangle$ is the average photon number, $n_c$ is the critical photon number required to saturate the TLS, $\beta$ is an empirical parameter describing TLS interaction, and $Q_{\mathrm{HP}}$ is the high-power quality factor.

Ring resonator losses are reported in Fig.~\ref{fig:ringresmeas}(e). Across eight resonances from four ring resonators on one chip, we extract a median TLS loss tangent of $(1.10 \pm 0.09) \times 10^{-6}$.

In Fig.~\ref{fig:ringresmeas}(e)-(h), we compare ring resonator values to those from more conventional $\lambda/4$ coplanar waveguide (CPW) resonators from the same wafer. $\lambda/4$ CPW resonators have center trace width (ground gap width) of 50~$\mu$m (25~$\mu$m), the same as the ring resonators. The $\lambda/4$ resonator chip is installed in a standard gold-plated copper microwave package, and wirebonded to a PCB for readout. This chip also has wirebonds across the on-chip feedline and across each resonator.

The median extracted $F\delta_{\mathrm{TLS}}^0$  values shown in Fig.~\ref{fig:ringresmeas}(e) agree for the two devices, as expected for chips from the same wafer with the same cross-sectional geometry. However, the ring resonators exhibit a smaller resonator-to-resonator spread in the extracted TLS loss tangent, with a standard deviation of $5.55 \times 10^{-8}$, compared to $1.52 \times 10^{-7}$ for the $\lambda/4$ devices. We attribute this to the factor of four difference in length between the two designs. Since the modes of a ring resonator average over a larger dielectric volume, they are less sensitive to local microscopic variations.~\cite{Chen2025} This is consistent with the ring  resonators' significantly higher median critical photon number $n_c$ (Fig.~\ref{fig:ringresmeas}(g)) - a device with a larger dielectric volume requires more input power to saturate a single TLS, as each photon will be averaged over a larger number of defects.

High-power quality factors $Q_{\mathrm{HP}}$  (Fig.~\ref{fig:ringresmeas}(f)) show an even starker difference in standard deviation between the two designs. Here, the median $Q_{\mathrm{HP}}$ is higher for the $\lambda/4$ devices in the standard package, but varies by four orders of magnitude between resonators on the same chip. This could be due to effects of wirebonds, PCB dielectrics, and normal metals in the package, creating a loss environment that is spatially inconsistent. Unlike conventional on-chip resonator setups in which the feedline and resonators are fabricated from the superconductors to ensure low-loss coupling in the high-power regime, the present high-throughput package relies on coupling through a normal-metal tungsten pin. A potential approach to address the suppressed $Q_{\mathrm{HP}}$ is therefore to replace the tungsten pin used in this work with a superconducting coupling element. This modification will be implemented in future work and is expected to significantly stabilize and enhance $Q_{\mathrm{HP}}$ while preserving the high-throughput functionality of the package.

Finally, the empirical parameter $\beta$  (Fig.~\ref{fig:ringresmeas}(h)) agrees between the two device sets. As this is generally considered a marker of TLS-TLS interaction strength,~\cite{Burnett2016} we expect $\beta$ to be the same for devices with the same cross-section and made of the same materials.

In conclusion, we have introduced a wirebond-free, PCB-free, drop-in microwave package designed for high-throughput characterization of on-chip superconducting devices. The package features a suspended tungsten transmission pin within a superconducting aluminum cavity, providing a clean microwave background with less than 3~dB ripple across the 4 GHz to 8~GHz band. By eliminating wirebonds and dielectric components inside the cavity, this design reduces unwanted microwave features and enables rapid chip swap-out.

Using this platform, we demonstrate precise measurements of superconducting ring resonators, extracting a TLS loss tangent of $(1.10 \pm 0.09) \times 10^{-6}$. Comparisons with $\lambda/4$ CPW resonators highlight the advantages of the ring geometry and packaging, confirming that our approach provides both compatible loss characterization and improved experimental efficiency. The ability to quickly measure many devices without compromising precision makes this system well-suited for generating large datasets to guide materials and device design improvements.

\begin{acknowledgments}

The contributions of W.-R.S., J.R.P., J.R., D.B., and C.R.H.M. are funded by the Materials Characterization and Quantum Performance: Correlation and Causation (MQC) program through collaboration between the Air Force Office of Scientific Research and the Laboratory for Physical Sciences under funding opportunity number FOA-AFRL-AFOSR-2023-0010.

A.M. and E.D. received funding through the University of Colorado Boulder College of Engineering and Applied Sciences as well as the Engineering Excellence Fund.

This work was conducted with the support of funding through the National Institute of Standards and Technology (NIST). Certain commercial materials and equipment are identified in this paper to foster understanding. Such identification does not imply recommendation or endorsement by NIST, nor does it imply that the materials or equipment identified is necessarily the best available for the purpose.

Thanks to Adam Sirois, Ben Mates, and Mark Keller for helpful feedback, and Nick Materise and David Pappas for thoughtful discussion.

\end{acknowledgments}

\bibliography{htcavity}

@article{Verjauw2021, title={Investigation of Microwave Loss Induced by Oxide Regrowth in High- Q Niobium Resonators}, volume={16}, ISSN={23317019}, DOI={10.1103/PhysRevApplied.16.014018}, note={arXiv: 2012.10761}, number={1}, journal={Physical Review Applied}, author={Verjauw, J. and Potočnik, A. and Mongillo, M. and Acharya, R. and Mohiyaddin, F. and Simion, G. and Pacco, A. and Ivanov, Ts and Wan, D. and Vanleenhove, A. and Souriau, L. and Jussot, J. and Thiam, A. and Swerts, J. and Piao, X. and Couet, S. and Heyns, M. and Govoreanu, B. and Radu, I.}, year={2021}, pages={1–14} }

@article{McRae2020, title={Materials loss measurements using superconducting microwave resonators}, volume={091101}, url={http://arxiv.org/abs/2006.04718}, DOI={10.1063/5.0017378}, number={September}, publisher={AIP Publishing, LLC}, author={McRae, Corey Rae Harrington and Wang, Haozhi and Gao, Jiansong and Vissers, Michael and Brecht, Teresa and Dunsworth, Andrew and Pappas, David and Mutus, Josh}, year={2020} }

@article{Pappas2011, title={Two level system loss in superconducting microwave resonators}, volume={21}, ISBN={1051-8223}, ISSN={10518223}, DOI={10.1109/TASC.2010.2097578}, note={Citation Key: Pappas2011}, number={3 PART 1}, journal={IEEE Transactions on Applied Superconductivity}, author={Pappas, David P. and Vissers, Michael R. and Wisbey, David S. and Kline, Jeffrey S. and Gao, Jiansong}, year={2011}, pages={871–874} }

@article{Ganjam2024, title={Surpassing millisecond coherence in on chip superconducting quantum memories by optimizing materials and circuit design}, volume={15}, ISSN={2041-1723}, DOI={10.1038/s41467-024-47857-6}, number={1}, journal={Nature Communications}, author={Ganjam, Suhas and Wang, Yanhao and Lu, Yao and Banerjee, Archan and Lei, Chan U and Krayzman, Lev and Kisslinger, Kim and Zhou, Chenyu and Li, Ruoshui and Jia, Yichen and Liu, Mingzhao and Frunzio, Luigi and Schoelkopf, Robert J.}, year={2024}, month=may, pages={3687}}

@article{Muller2019, title={Towards understanding two-level-systems in amorphous solids - Insights from quantum circuits}, volume={82},  number={124501}, journal={Rep. Prog. Phys.}, author={Müller, Clemens and Cole, Jared H. and Lisenfeld, Jürgen}, year={2019} }

@article{Baity2024, title={Circle fit optimization for resonator quality factor measurements: Point redistribution for maximal accuracy}, volume={6}, rights={https://creativecommons.org/licenses/by/4.0/}, ISSN={2643-1564}, url={https://link.aps.org/doi/10.1103/PhysRevResearch.6.013329}, DOI={10.1103/physrevresearch.6.013329}, number={1}, journal={Physical Review Research}, publisher={American Physical Society (APS)}, author={Baity, Paul G. and Maclean, Connor and Seferai, Valentino and Bronstein, Joe and Shu, Yi and Hemakumara, Tania and Weides, Martin}, year={2024}, month=mar}

@article{Woods2019, title={Determining Interface Dielectric Losses in Superconducting Coplanar-Waveguide Resonators}, volume={12}, ISSN={2331-7019}, DOI={10.1103/physrevapplied.12.014012},  note={Citation Key: Woods2019}, number={1}, journal={Physical Review Applied}, publisher={American Physical Society}, author={Woods, W. and Calusine, G. and Melville, A. and Sevi, A. and Golden, E. and Kim, D.K. and Rosenberg, D. and Yoder, J.L. and Oliver, W.D.}, year={2019}, pages={1} }

@article{Bal2024, title={Systematic improvements in transmon qubit coherence enabled by niobium surface encapsulation}, volume={10}, rights={2024 This is a U.S. Government work and not under copyright protection in the US; foreign copyright protection may apply}, ISSN={2056-6387}, DOI={10.1038/s41534-024-00840-x}, number={1}, journal={npj Quantum Information}, publisher={Nature Publishing Group}, author={Bal, Mustafa and Murthy, Akshay A. and Zhu, Shaojiang and Crisa, Francesco and You, Xinyuan and Huang, Ziwen and Roy, Tanay and Lee, Jaeyel and Zanten, David van and Pilipenko, Roman and Nekrashevich, Ivan and Lunin, Andrei and Bafia, Daniel and Krasnikova, Yulia and Kopas, Cameron J. and Lachman, Ella O. and Miller, Duncan and Mutus, Josh Y. and Reagor, Matthew J. and Cansizoglu, Hilal and Marshall, Jayss and Pappas, David P. and Vu, Kim and Yadavalli, Kameshwar and Oh, Jin-Su and Zhou, Lin and Kramer, Matthew J. and Lecocq, Florent and Goronzy, Dominic P. and Torres-Castanedo, Carlos G. and Pritchard, P. Graham and Dravid, Vinayak P. and Rondinelli, James M. and Bedzyk, Michael J. and Hersam, Mark C. and Zasadzinski, John and Koch, Jens and Sauls, James A. and Romanenko, Alexander and Grassellino, Anna}, year={2024}, month=apr, pages={1–8}}

@article{Rieger2023, title={Fano Interference in Microwave Resonator Measurements}, volume={20}, ISSN={2331-7019}, DOI={10.1103/PhysRevApplied.20.014059}, number={1}, journal={Physical Review Applied}, author={Rieger, D. and Günzler, S. and Spiecker, M. and Nambisan, A. and Wernsdorfer, W. and Pop, I.M.}, year={2023}, month=july, pages={014059}}

@article{Khalil2012, title={An analysis method for asymmetric resonator transmission applied to superconducting devices}, volume={111}, ISSN={00218979}, DOI={10.1063/1.3692073}, note={arXiv: 1108.3117
Citation Key: Khalil2012}, number={5}, journal={Journal of Applied Physics}, author={Khalil, M. S. and Stoutimore, M. J.A. and Wellstood, F. C. and Osborn, K. D.}, year={2012} }

@article{Read2023, title={Precision Measurement of the Microwave Dielectric Loss of Sapphire in the Quantum Regime with Parts-per-Billion Sensitivity}, volume={19}, ISSN={2331-7019}, DOI={10.1103/PhysRevApplied.19.034064}, number={3}, journal={Physical Review Applied}, author={Read, Alexander P. and Chapman, Benjamin J. and Lei, Chan U and Curtis, Jacob C. and Ganjam, Suhas and Krayzman, Lev and Frunzio, Luigi and Schoelkopf, Robert J.}, year={2023}, month=mar, pages={034064}}

@article{Pitten2025, title={Effective Reflection Mode Measurement for Hanger-Coupled Microwave Resonators}, author={Pitten, John R. and Materise, Nicholas and Syong, Wei-Ren and Ramirez, Jorge and Bennett, Douglas and McRae, Corey Rae H.}, journal={Phys. Rev. A}, volume={112}, issue={5}, pages={052618}, numpages={14}, year={2025}, month={Nov}, publisher={American Physical Society}, doi={10.1103/71lw-rsrj}, url={https://link.aps.org/doi/10.1103/71lw-rsrj}}

@article{Bronn2018, title={High Coherence Plane Breaking Packaging for Superconducting Qubits}, publisher={IOP Publishing}, author={Bronn, Nicholas T and Adiga, Vivekananda P and Olivadese, Salvatore B and Wu, Xian and Chow, Jerry M and Pappas, David P}, year={2018} }

@article{Wenner2011, title={Wirebond crosstalk and cavity modes in large chip mounts for superconducting qubits}, volume={24}, ISSN={0953-2048}, DOI={10.1088/0953-2048/24/6/065001}, note={arXiv: 1011.4982
Citation Key: Wenner2011
ISBN: 0953-2048}, number={6}, journal={Superconductor Science and Technology}, author={Wenner, J and Neeley, M and Bialczak, Radoslaw C and Lenander, M and Lucero, Erik and O’Connell, A D and Sank, D and Wang, H and Weides, M and Cleland, A. N. and Martinis, John M.}, year={2011}, pages={065001} }

@article{McConkey2018, title={Mitigating leakage errors due to cavity modes in a superconducting quantum computer}, volume={3}, ISSN={2058-9565}, DOI={10.1088/2058-9565/aabd41}, number={3}, journal={Quantum Science and Technology}, author={McConkey, T G and Béjanin, J H and Earnest, C T and McRae, C R H and Pagel, Z and Rinehart, J R and Mariantoni, M}, year={2018}, month=july, pages={034004}}

@article{McRae2017, title={Thermocompression Bonding Technology for Multilayer Superconducting Quantum Circuits}, volume={111}, ISSN={0003-6951}, url={https://arxiv.org/abs/1705.02435}, DOI={10.1063/1.5003169}, note={arXiv: 1705.02435
Citation Key: McRae2017}, number={123501}, journal={Applied Physics Letters}, author={McRae, C. R. H. and Béjanin, J. H. and Pagel, Z. and Abdallah, A. O. and McConkey, T. G. and Earnest, C. T. and Rinehart, J. R. and Mariantoni, M.}, year={2017} }

@article{Place2021, title={New material platform for superconducting transmon qubits with coherence times exceeding 0.3 milliseconds}, volume={12}, ISSN={20411723}, url={http://dx.doi.org/10.1038/s41467-021-22030-5}, DOI={10.1038/s41467-021-22030-5}, number={1}, journal={Nature Communications}, author={Place, Alexander P.M. and Rodgers, Lila V.H. and Mundada, Pranav and Smitham, Basil M. and Fitzpatrick, Mattias and Leng, Zhaoqi and Premkumar, Anjali and Bryon, Jacob and Vrajitoarea, Andrei and Sussman, Sara and Cheng, Guangming and Madhavan, Trisha and Babla, Harshvardhan K. and Le, Xuan Hoang and Gang, Youqi and Jäck, Berthold and Gyenis, András and Yao, Nan and Cava, Robert J. and de Leon, Nathalie P. and Houck, Andrew A.}, year={2021} }

@article{Chen2025, title={Efficient methods for extracting superconducting resonator loss in the single-photon regime}, volume={137}, ISSN={0021-8979, 1089-7550}, DOI={10.1063/5.0242201}, number={4}, journal={Journal of Applied Physics}, author={Chen, Cliff and Perello, David and Aghaeimeibodi, Shahriar and Marcaud, Guillaume and Jarrige, Ignace and Lee, Hanho and Fon, Warren and Matheny, Matt and Gao, Jiansong}, year={2025}, month=jan, pages={044401}}

@book{Pozar,
  author    = {David M. Pozar},
  title     = {Microwave Engineering},
  publisher = {John Wiley \& Sons},
  year      = {2012},
  edition   = {4th},
  isbn      = {978-0-470-63155-3}
}

@article{Sun2025,
  title        = {Superconducting Ring Resonators: Modelling, Simulation, and Experimental Characterisation},
  author       = {Sun, Zhenyuan and Withington, Stafford and Thomas, Christopher and Zhao, Songyuan},
  journal      = {arXiv preprint arXiv:2506.23811},
  year         = {2025},
  url          = {https://arxiv.org/abs/2506.23811}
}

@phdthesis{GaoThesis, title={The Physics of Superconducting Microwave Resonators}, volume={2008}, url={http://resolver.caltech.edu/CaltechETD:etd-06092008-235549}, DOI={10.1088/0031-9120/1/1/306}, author={Gao, Jiansong}, year={2008} }

@article{Burnett2016, title={Analysis of high quality superconducting resonators: consequences for TLS properties in amorphous oxides}, volume={29}, journal={Supercond. Sci. Technol}, publisher={IOP Publishing}, author={Burnett, J and Faoro, L and Lindström, T}, year={2016} }

\end{document}



\title{Supplementary Material:\\High-Throughput Microwave Package for \\Precise Superconducting Device Measurement} 



\author{Wei-Ren Syong}
\affiliation{Department of Physics, University of Colorado, Boulder, CO 80309, USA}
\affiliation{Department of Electrical, Computer, and Energy Engineering, University of Colorado, Boulder, CO 80309, USA}
\affiliation{National Institute of Standards and Technology, Boulder, CO 80305, USA}
\author{Allie Miller}
\affiliation{Department of Electrical, Computer, and Energy Engineering, University of Colorado, Boulder, CO 80309, USA}
\author{Emma Davis}
\affiliation{Department of Mechanical Engineering, University of Colorado, Boulder, CO 80309, USA}
\author{John R. Pitten}
\affiliation{Department of Physics, University of Colorado, Boulder, CO 80309, USA}
\affiliation{Department of Electrical, Computer, and Energy Engineering, University of Colorado, Boulder, CO 80309, USA}
\affiliation{National Institute of Standards and Technology, Boulder, CO 80305, USA}
\author{Jorge Ramirez}
\affiliation{Department of Physics, University of Colorado, Boulder, CO 80309, USA}
\affiliation{Department of Electrical, Computer, and Energy Engineering, University of Colorado, Boulder, CO 80309, USA}
\affiliation{National Institute of Standards and Technology, Boulder, CO 80305, USA}
\author{Nathan Ortiz}
\affiliation{National Institute of Standards and Technology, Boulder, CO 80305, USA}
\author{Michael Vissers}
\affiliation{National Institute of Standards and Technology, Boulder, CO 80305, USA}
\author{Doug Bennett}
\affiliation{National Institute of Standards and Technology, Boulder, CO 80305, USA}
\author{Corey~Rae Harrington McRae}
\affiliation{Department of Electrical, Computer, and Energy Engineering, University of Colorado, Boulder, CO 80309, USA}
\affiliation{Department of Physics, University of Colorado, Boulder, CO 80309, USA}
\affiliation{National Institute of Standards and Technology, Boulder, CO 80305, USA}

\date{\today}

\pacs{}

\maketitle 

\section{Additional cavity package design details}

The sample package including SMP holders has external dimensions of 29~mm (length) $\times$ 19~mm (width) $\times$ 18~mm (height). Each chip is 7.5~mm $\times$ 7.5~mm and is placed in a recessed pocket in the package base. Since the chip is held in place with pogo pins, many chip thicknesses can be accommodated. The package CAD design and component list is available on GitHub.~\footnote{Designs hosted at \hyperref[https://github.com/Boulder-Cryogenic-Quantum-Testbed/highthroughput-cavity-package]{https://github.com/Boulder-Cryogenic-Quantum-Testbed/highthroughput-cavity-package}}.

The cavity interior can be separated into three functional regions that control the electromagnetic field environment from the readout pin to the chip. The tungsten pin is positioned within a rectangular cutout below the main cavity, forming a coax-like geometry whose characteristic impedance in $\Omega$ can be estimated using the coaxial waveguide approximation:
\begin{equation}
    Z_0 = \frac{138}{\sqrt{\varepsilon_\mathrm{r}}} \ln\!\left(\frac{D}{d}\right),
    \label{eq:coaximpedance}
\end{equation}
where $d$ is the pin diameter and $D$ is the effective diameter defined by the rectangular cutout. The cutout edge length of approximately 1~mm maintains an effective impedance of $Z_0 \approx 50~\Omega$, minimizing reflections.

Above the pin pocket, a horn-like transition gradually expands the cross-sectional area between the pin pocket and the larger chip cavity.

The top region provides space for the chip. Two narrow ledges support the chip, while non-magnetic pogo pins press the chip against the seating surface to fix its position relative to the W pin without requiring paste or clamps. The pogo pins are placed on the opposite side of the coupling region in order to minimize perturbations to the local field distribution and the coupling between the chip and the pin.

\begin{figure}
\includegraphics[width=0.6\linewidth]{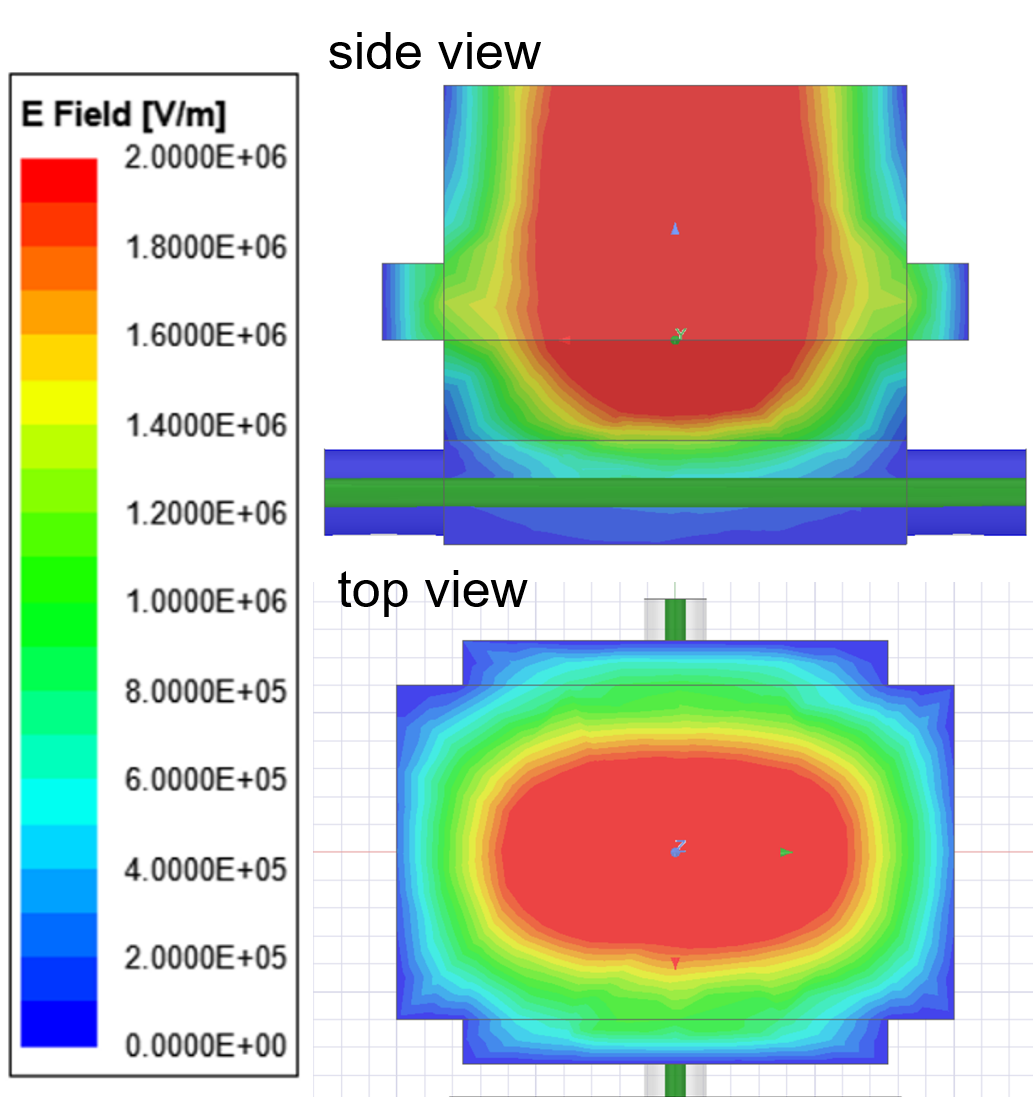}%
\caption{Simulated electric-field magnitude of the cavity at its fundamental resonance frequency which is near 28~GHz.
\label{fig:cavity_efield}}%
\end{figure}

The simulated electric field distribution of the fundamental cavity mode near 28 GHz is shown in Fig.~\ref{fig:cavity_efield}.

\section{Additional ring resonator design details}

\begin{figure}
\includegraphics[width=1.0\linewidth]{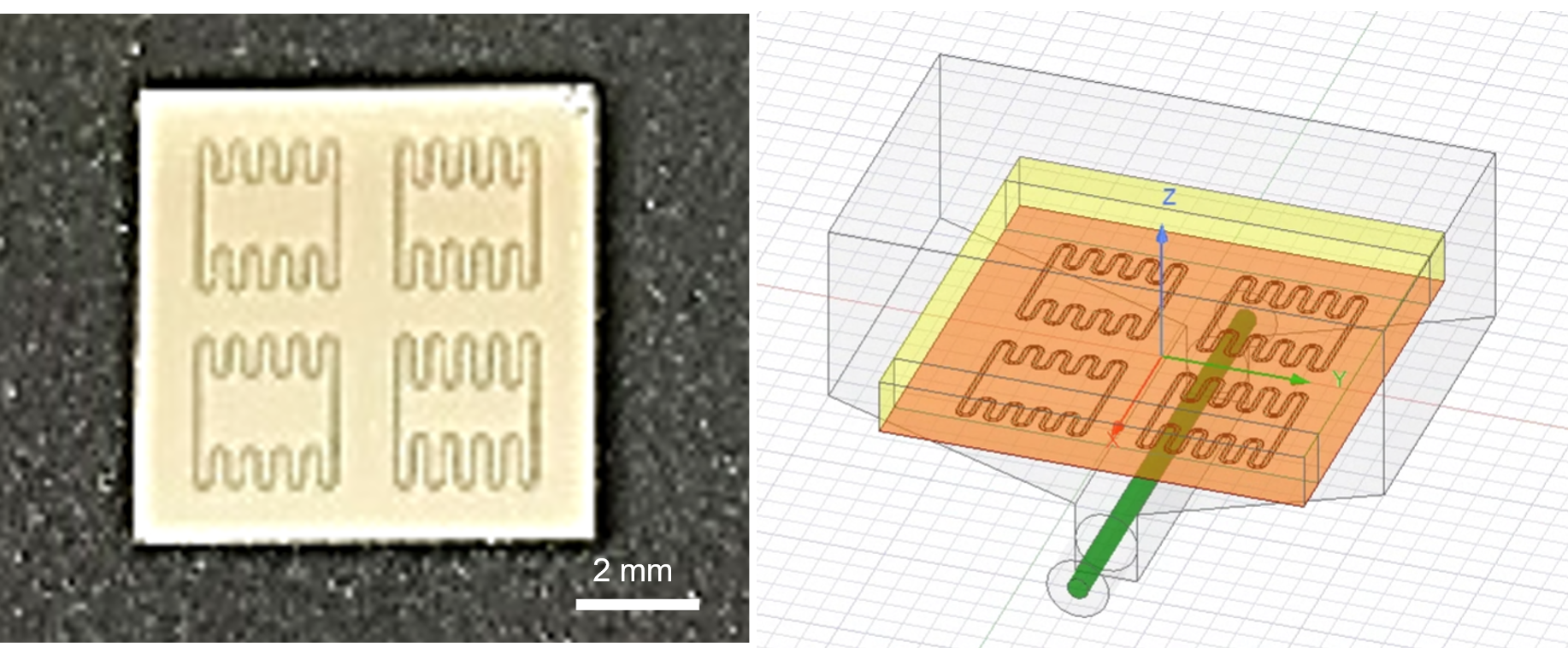}%
\caption{Ring resonator chip. Left: Photograph of fabricated chip. Right: Ansys HFSS image of chip in cavity, with transmission pin in green.}
\label{fig:chipimage}%
\end{figure}

Fig.~\ref{fig:chipimage} shows a full image of the fabricated ring resonator chip. Each ring resonator has a slightly different perimeter, resulting in a distinct resonance frequency within the 4-8~GHz range and allowing multiplexed readout through the transmission pin. 

The perimeter, simulated resonance frequency using Ansys HFSS, experimentally measured resonance frequency, resonance frequency shift ($\alpha$), and mode splitting ($\Delta f_\mathrm{c}$) for each ring are reported in Tab.~\ref{tab:tableringsimmea}. Here, the resonance frequency shift is defined as $\alpha \equiv f_\mathrm{c}^\mathrm{meas} - f_\mathrm{c}^\mathrm{sim}$. Across all devices, the measured resonance frequencies are systematically lower than the simulated values, resulting in a consistent negative frequency shift. Following Ref.~\citenum{Sun2025}, this shift is attributed to a reduction in the effective phase velocity due to distributed electromagnetic loading along the resonator. In contrast, the observed mode splitting remains small and varies weakly between resonators, consistent with frequency splitting arising from impedance perturbations associated with fabrication-induced non-uniformities. The observed behavior agrees with the theoretical analysis and equations presented in Ref.~\citenum{Sun2025}.

Terminated CPW resonators are susceptible to slotline modes, which occur when there is an imbalance in the ground-plane potentials on either side of the center conductor.~\cite{Pozar} Slotline modes can be near-resonant with on-chip devices, complicating data interpretation. In our design, no slotline-mode resonances were observed in measurement. HFSS simulations further confirm that the lowest-order slotline modes are strongly detuned from the operational frequency range. As shown in Tab.~\ref{tab:tableringsimmea}, the fundamental CPW modes for all four rings lie within the 4-8 GHz band, while the lowest-order slotline modes occur above 10 GHz. 

\begin{figure}
\includegraphics[width=1.0\linewidth]{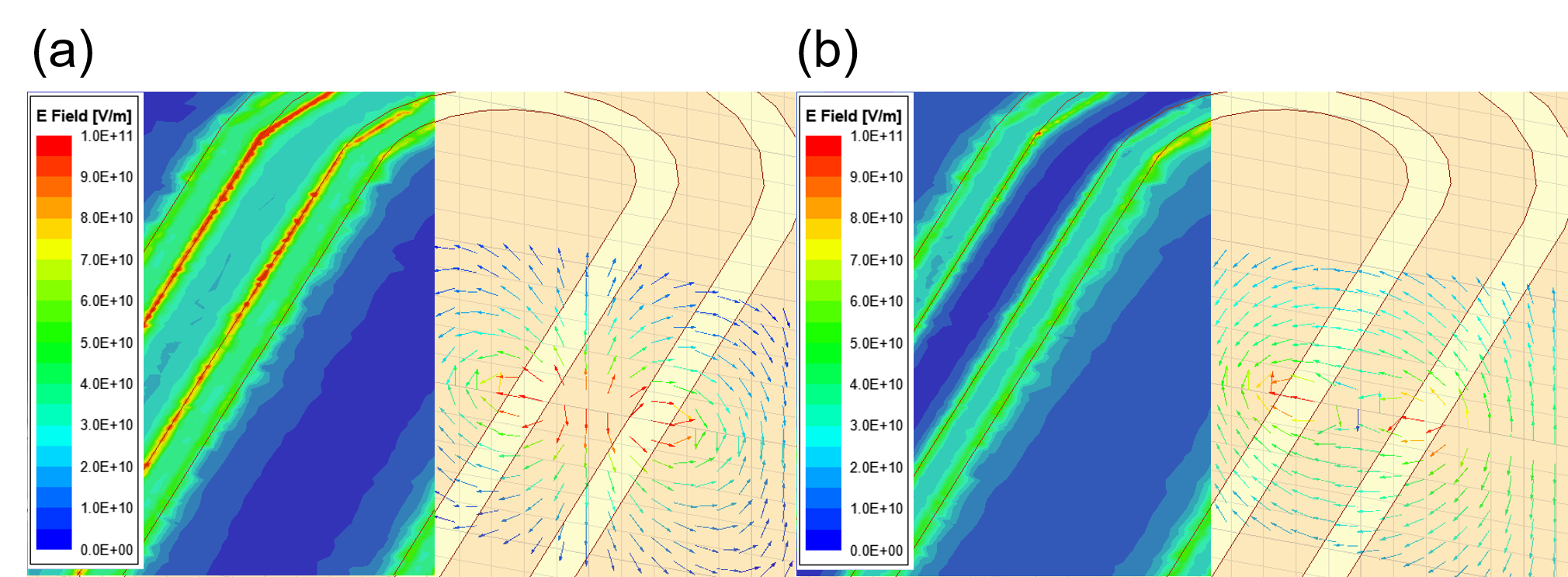}%
\caption{Simulated electric-field magnitude and direction of the ring resonator, using Resonator~0 as an example. (a) CPW mode and (b) slotline mode.}  
\label{fig:Efieldcpwslotlinesim}%
\end{figure}

The distinction between CPW and slotline modes can be observed by examining their electric field and the field direction. As shown in Fig.~\ref{fig:Efieldcpwslotlinesim}, the slotline modes are characterized by strongly concentrated E-fields within the CPW gaps, with the fields in the two grounds oriented in opposite directions. In contrast, the CPW modes exhibit more uniform field distributions across the signal and ground regions, with the fields on both sides of the grounds oriented in the same direction.

\begin{table}[]
    \centering
    \begin{tabular}{c c c c c}
        \textbf{Resonator} & \textbf{0} & \textbf{1} & \textbf{2} & \textbf{3} \\
        \hline
        Perimeter (mm)                 & 20.4126 & 19.1326 & 18.1246 & 17.3246 \\
        Simulated $f_\mathrm{c}$ (GHz) &  6.5463 &  6.9413 &  7.3295 &  7.7028 \\
        Measured $f_\mathrm{c}$ (GHz)  &  6.072 &  6.474 &  6.829 &  7.139 \\
        Simulated slotline  $f_\mathrm{0}$ (GHz) & 11.2102 & 11.6601 & 12.0954 & 12.5030 \\
        $\alpha$ (GHz)                 & -0.4744 & -0.4677 & -0.5009 & -0.5642 \\
        $\Delta f_\mathrm{c}$ (GHz)    &  0.0063 &  0.0042 &  0.0028 &  0.0021 \\
    \end{tabular}
    \caption{Summary of designed ring resonators. For each resonator, we list the perimeter, HFSS-simulated resonance frequency, measured resonance frequency, resonance-frequency shift ($\alpha$), and mode splitting ($\Delta f_\mathrm{c}$). The $\alpha$ values are estimated using the average of the split modes to reduce the effect of impedance perturbations.}
    \label{tab:tableringsimmea}
\end{table}

\section{HFSS simulation details}

\begin{figure}
\includegraphics[width=0.9\linewidth]{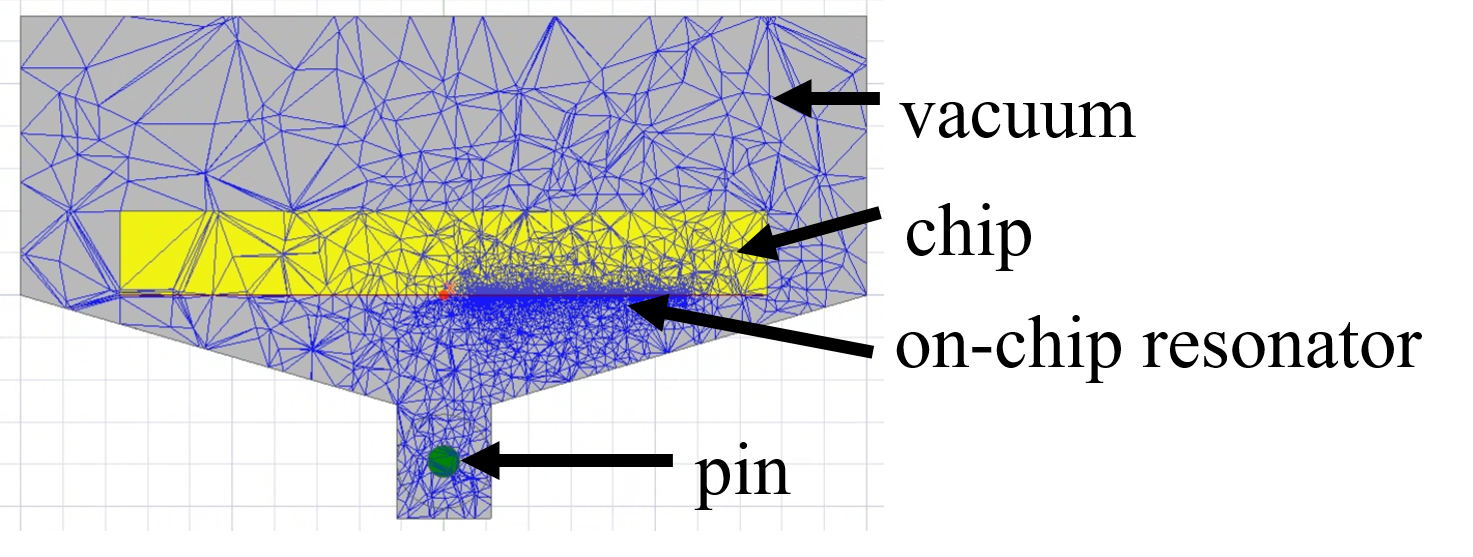}%
\caption{HFSS simulation mesh shown in a side view of the cavity. Mesh elements are displayed in blue and are refined to be denser in regions with strong electric fields, such as near the resonator and the W transmission pin.
\label{fig:meshing}}%
\end{figure}

Electromagnetic simulations were performed using Ansys HFSS and Maxwell to optimize the cavity geometry, estimate resonance frequencies, and calculate participation ratios. An adaptive meshing process was employed to ensure convergence of both the simulated $S_{21}$ response and the electromagnetic field distributions. The mesh density was refined in regions of strong electric field, particularly near the ring resonator and the suspended transmission pin, as shown in Fig.~\ref{fig:meshing}. This localized mesh refinement improves accuracy of the results and the computation time.

\section{Coupling simulations}

\begin{figure}
\includegraphics[width=0.9\linewidth]{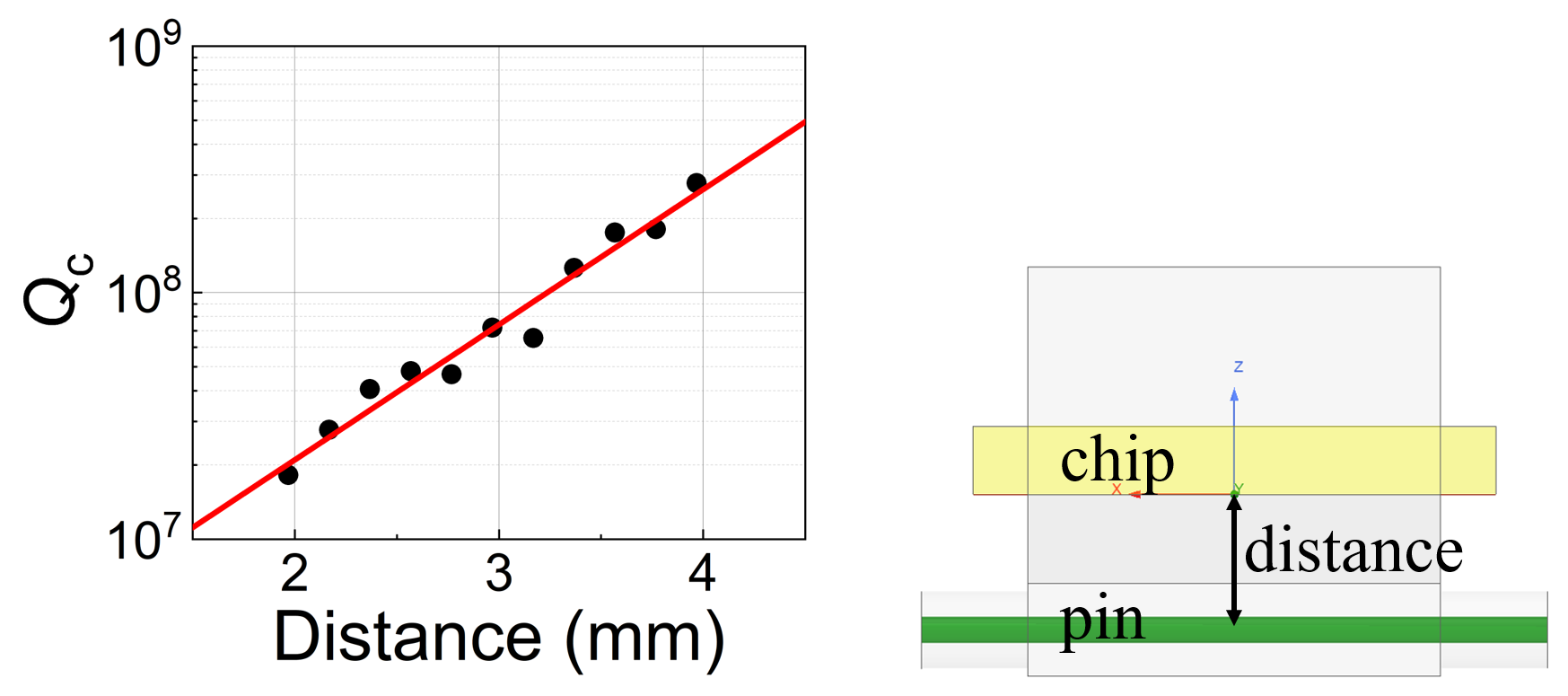}%
\caption{HFSS simulation of $Q_c$ as a function of the distance between the on-chip resonator and the transmission pin. Red line represents a fit to simulation results, demonstrating that $Q_c$ increases predictably with increasing pin-to-chip distance.}
\label{fig:couplingsim}%
\end{figure}

The coupling between the transmission pin and the on-chip resonator is simulated to determine the dependence of the external coupling quality factor ($Q_c$) on the pin-to-chip distance. These simulations can be used to guide the design of multiple package bases with varying chip heights, allowing precise tuning of $Q_c$ without requiring any modifications to the on-chip resonator geometry. This modular approach is critical for optimizing the readout coupling strength for individual experiments, especially  when the on-chip resonator $Q_i$ is not  well-known \textit{a priori}. As shown in Fig.~\ref{fig:couplingsim}, $Q_c$ increases with pin-to-chip distance.

\section{Clean microwave background of the cavity}

\begin{figure}
\includegraphics[width=1.0\linewidth]{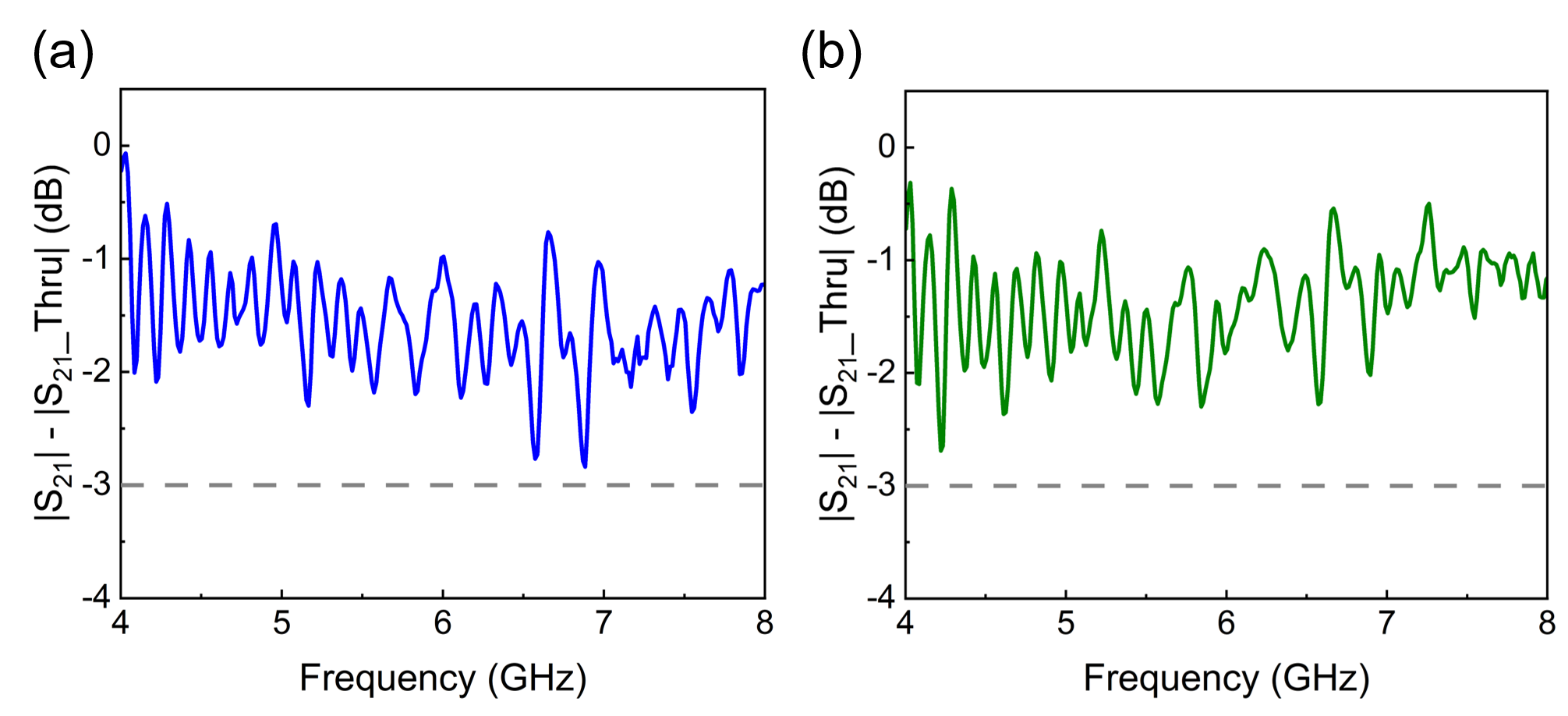}%
\caption{Difference in $|S_{21}|$ of the cavity and associated thru, (a) without and (b) with the inserted chip.
\label{fig:caavitymearemovebg}}%
\end{figure}

Fig.~\ref{fig:caavitymearemovebg} shows the broadband transmission background of the cavity after subtracting the magnitude response of a reference through. Fig.~\ref{fig:caavitymearemovebg}(a) and (b) correspond to measurements performed with an empty cavity (cooldown 1) and with the cavity loaded with the resonator chip (cooldown 2), respectively. In both cases, the background transmission remains within 3~dB across the operating range from 4–8~GHz. This indicates that the cavity introduces only a small and nearly chip-independent insertion loss over the measurement band. Therefore, resonator features can be reliably characterized without additional background artifacts introduced by the cavity itself.

\begin{figure}
\includegraphics[width=1.0\linewidth]{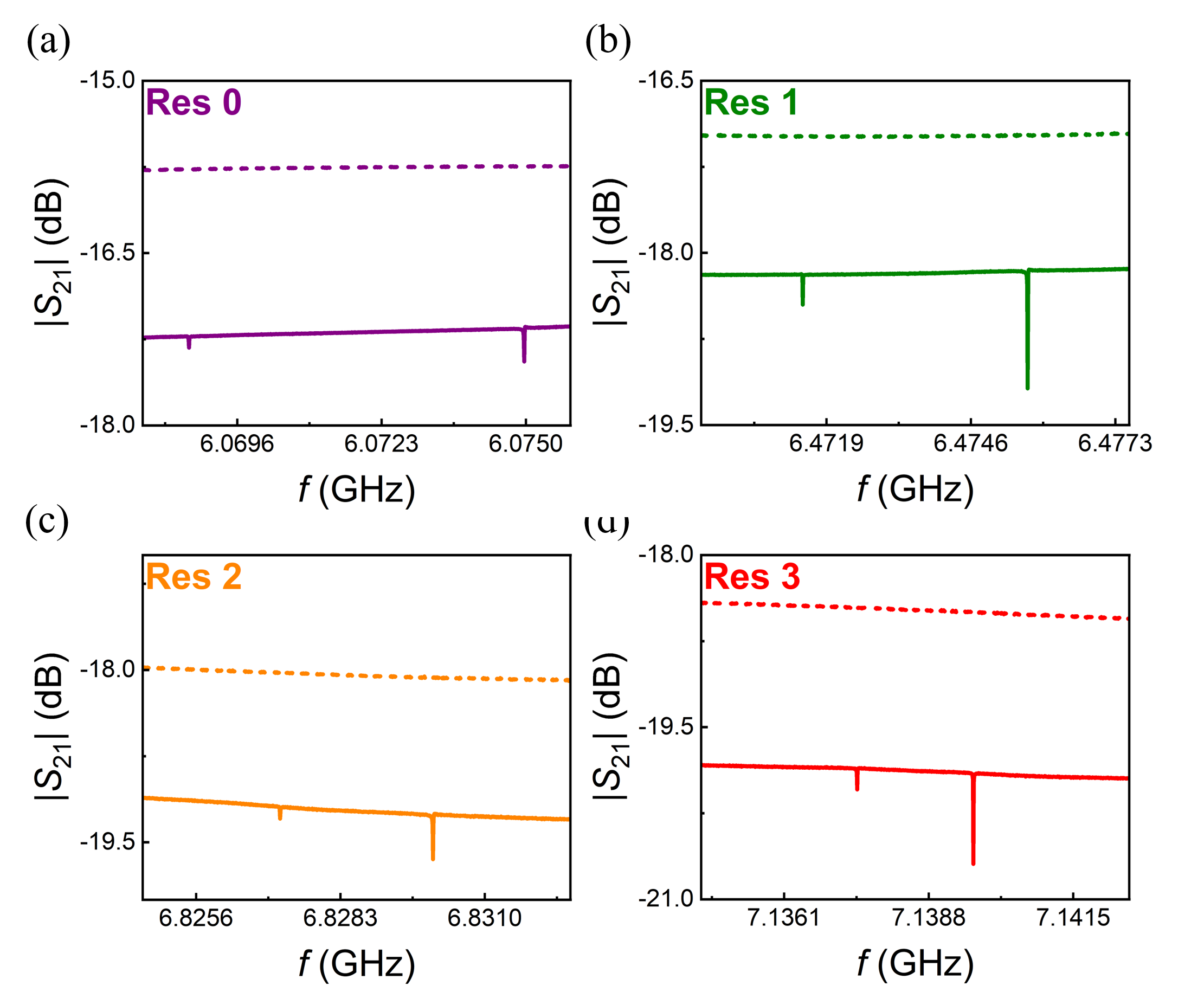}%
\caption{Transmission background around representative resonances, compared with the corresponding through response. Each panel shows a limited frequency span covering the split resonance dips of each ring resonator.}
\label{fig:caavitynarrowbg}
\end{figure}

To further assess the impact of the cavity background on resonator measurements, we examine the transmission in the frequency spans around individual resonances. Fig.~\ref{fig:caavitynarrowbg} compares the measured $S_{21}$ background in the vicinity of several representative resonances to the corresponding through. Over these spans, the background exhibits weak attenuation and negligible nonlinear frequency dependence.

\section{Wiring diagram}

\begin{figure}
\includegraphics[width=1.0\linewidth]{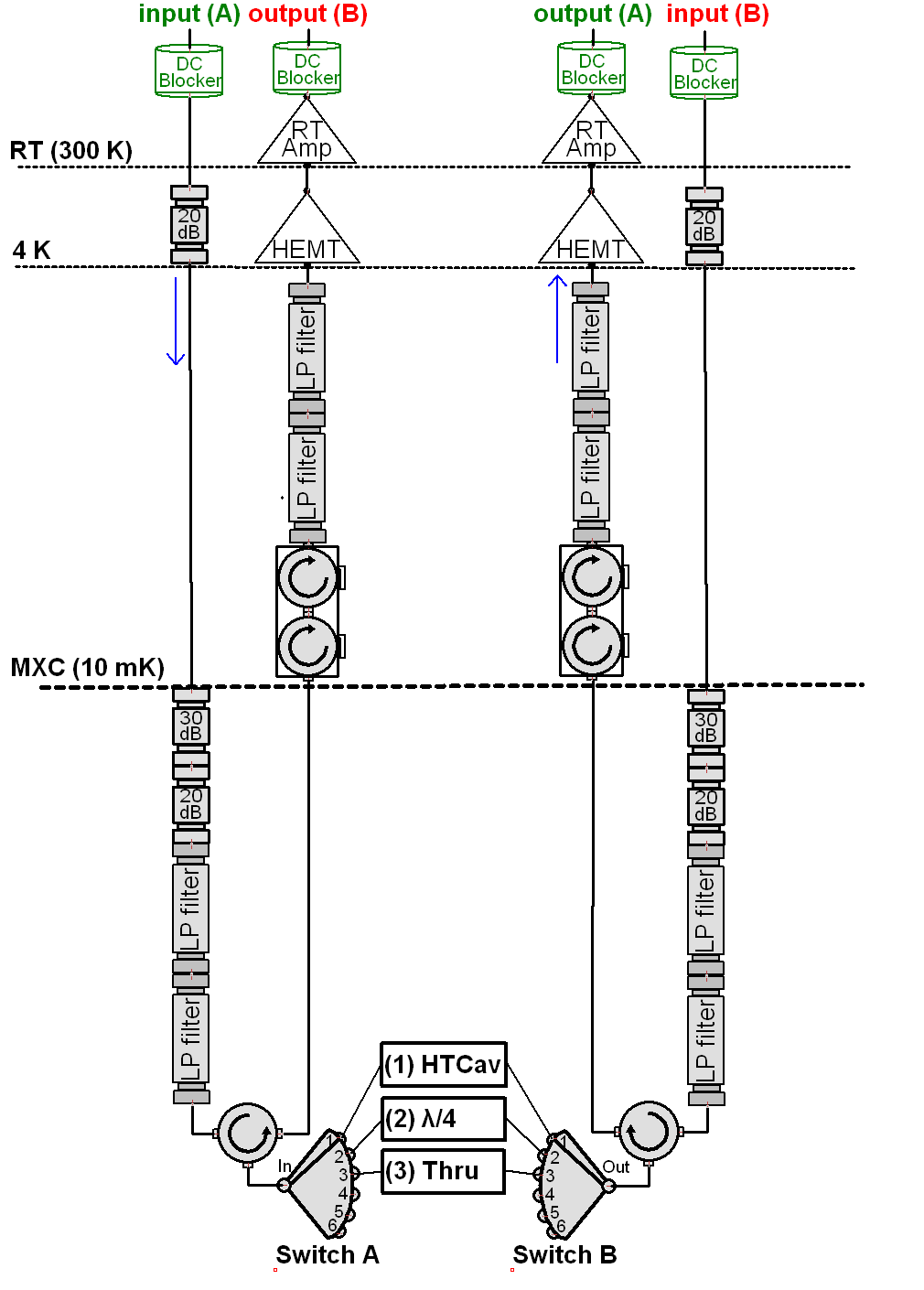}%
\caption{Wiring diagram. The schematic shows the distributed RF components, cryogenic and room-temperature amplifiers, and the two six-way cryogenic switches. Connections to the sample package that is mounted at the mixing chamber stage of the dilution refrigerator are also illustrated. This configuration enables rapid, reproducible measurements of multiple devices without warming up the dilution refrigerator.
\label{fig:wiringdiagram}}
\end{figure}

The complete cryogenic wiring diagram for the experiment is shown in Fig.~\ref{fig:wiringdiagram}. The input microwave lines include a total of 70~dB of attenuation, distributed between the 4~K stage and the MC stage. On the output side, signals are first amplified by a high-electron-mobility transistor (HEMT) amplifier mounted at the 4~K stage, followed by a low-noise room-temperature amplifier located outside the DR. This two-stage amplification chain ensures sufficient gain for sensitive measurements.

Two six-way cryogenic microwave switches are integrated into the input and output lines, allowing the switch between multiple devices in situ. This feature enables direct comparisons between different resonator chips under identical measurement conditions.

\begin{figure}
\includegraphics[width=0.8\linewidth]{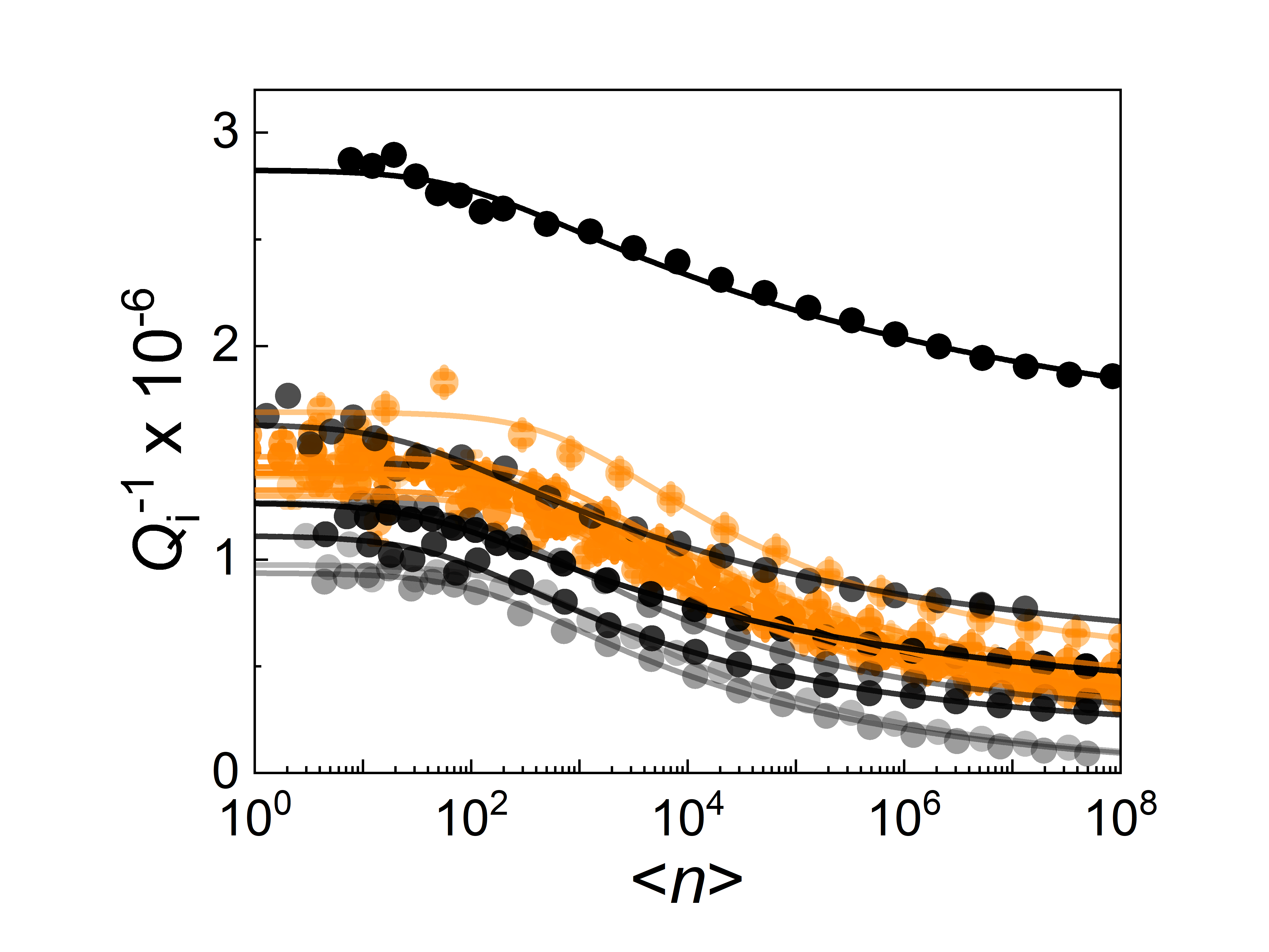}%
\caption{Measured $Q_i^{-1}$. The black data correspond to $\lambda/4$ resonators installed in a conventional package, while the orange data correspond to the ring resonators measured using our high-throughput cavity.
\label{fig:Qiinverse}
}%
\end{figure}

\section{Measured $Q_i^{-1}$ of the resonators in conventional packaging}

Fig.~\ref{fig:Qiinverse} shows the measured $Q_i^{-1}$ as a function of the average photon number $\langle n \rangle$ for resonators measured in two different platforms. The black data correspond to $\lambda/4$ resonators installed in a conventional package, while the orange data correspond to the ring resonators measured using our high-throughput cavity which are also shown in Fig. 4 in the main text. The $\lambda/4$ resonators are designed to have the same coplanar waveguide cross-sectional geometry (center trace width, gap, and film stack) as the ring resonators, enabling a direct comparison between the two device.

Across the power range, the $\lambda/4$ resonators in the conventional package exhibit larger device-to-device variation in $Q_i^{-1}$. This spread could be due to wire bonds, PCB materials, and other lossy elements required in standard packaging.

However, the ring resonators measured in our cavity show a narrower distribution of $Q_i^{-1}$. The reduced spread demonstrates the uniform and low-loss environment provided by the high-throughput cavity. These results support the conclusion that the cavity platform yields more reproducible and reliable measurements of intrinsic resonator performance.

\bibliography{htcavity}